\documentclass[twocolumn]{aa}
\usepackage{graphicx}
\usepackage{txfonts}

\begin{document}
\title{Timescale of variation and the size of the accretion
 disc in active galactic nuclei}

\author{M.R.S. Hawkins\inst{1}}

\offprints{M.R.S. Hawkins}

\institute{Institute for Astronomy (IfA), University of Edinburgh,
              Royal Observatory, Blackford Hill, Edinburgh EH9 3HJ\\
              \email{mrsh@roe.ac.uk}
             }

\date{Received November ??, 2004; accepted ??}

\abstract{
This paper sets out to measure the timescale of quasar variability with
a view to new understanding of the size of accretion discs in active
galactic nuclei.  Previous attempts to measure such timescales have
been based on sparsely sampled data covering small ranges of time.
Here we combine data from two large scale monitoring programmes to
obtain Fourier power spectra of light curves covering nearly three
orders of magnitude in frequency in blue and red passbands.  If the
variations are interpreted as due to gravitational microlensing, then
timescale measurements in the observer's frame imply a minimum mass for
the microlensing bodies of around $0.4 M_{\odot}$.  On the
assumption that the variations are intrinsic to the quasars, a
correction must be made for time dilation.  In this case the power
spectrum shows a break corresponding to a timescale of about 11 years.
This timescale is used to measure the size of the accretion disc, which
is found to be about $10^{-2}$ pc or 10 light days, in agreement with
limits set by self-gravitation and coincident with the broad line
region of the active galactic nucleus.  It is suggested that the broad
line region may be associated with the break up of the outer part of
the accretion disc.

   }

\maketitle
%

\begin{figure*}[t]
\setlength{\unitlength}{1mm}
\begin{picture} (200,40) (0,100) 
\includegraphics[width=0.5\textwidth]{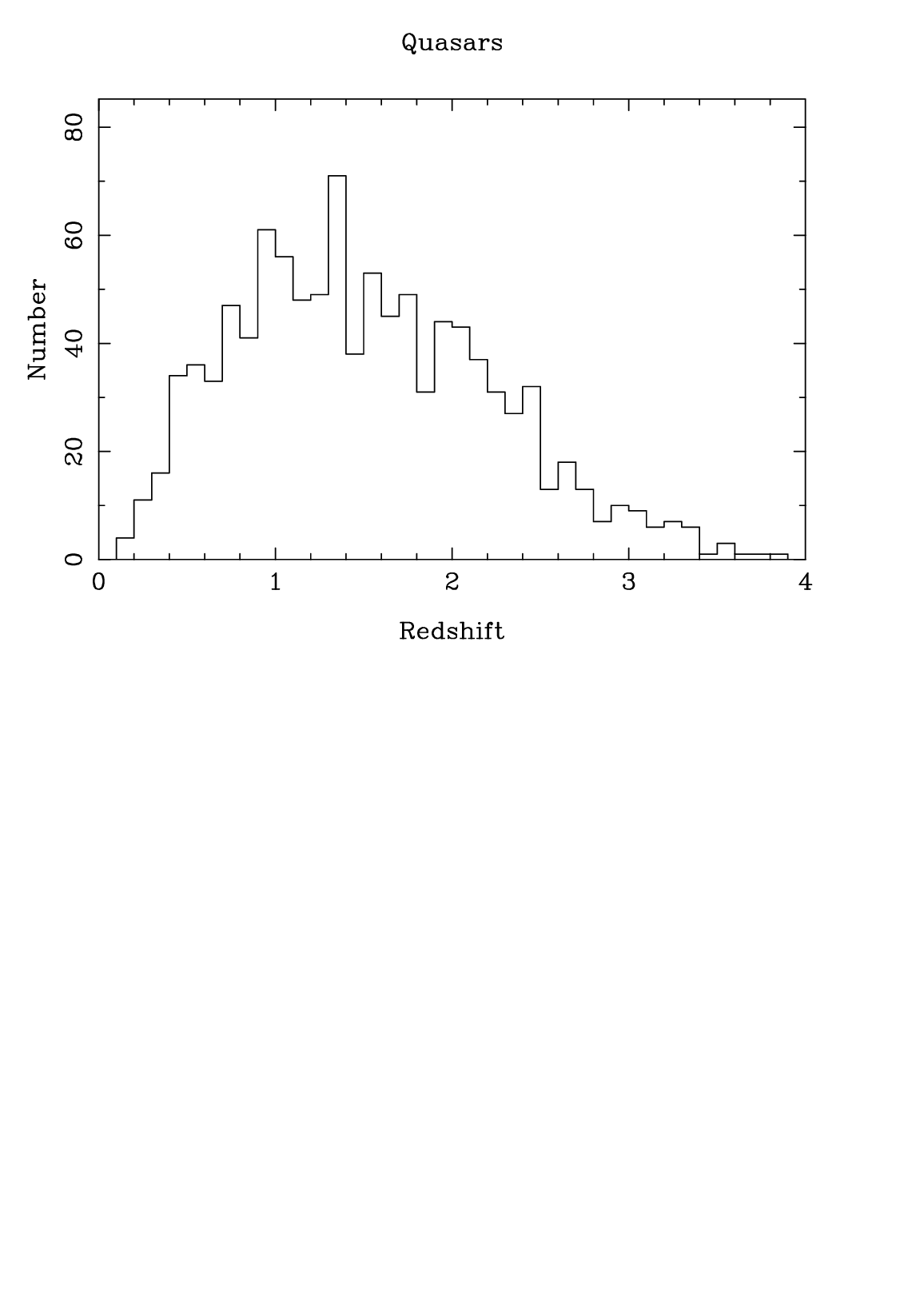}
\end{picture}
\begin{picture} (200,40) (-90,60) 
\includegraphics[width=0.5\textwidth]{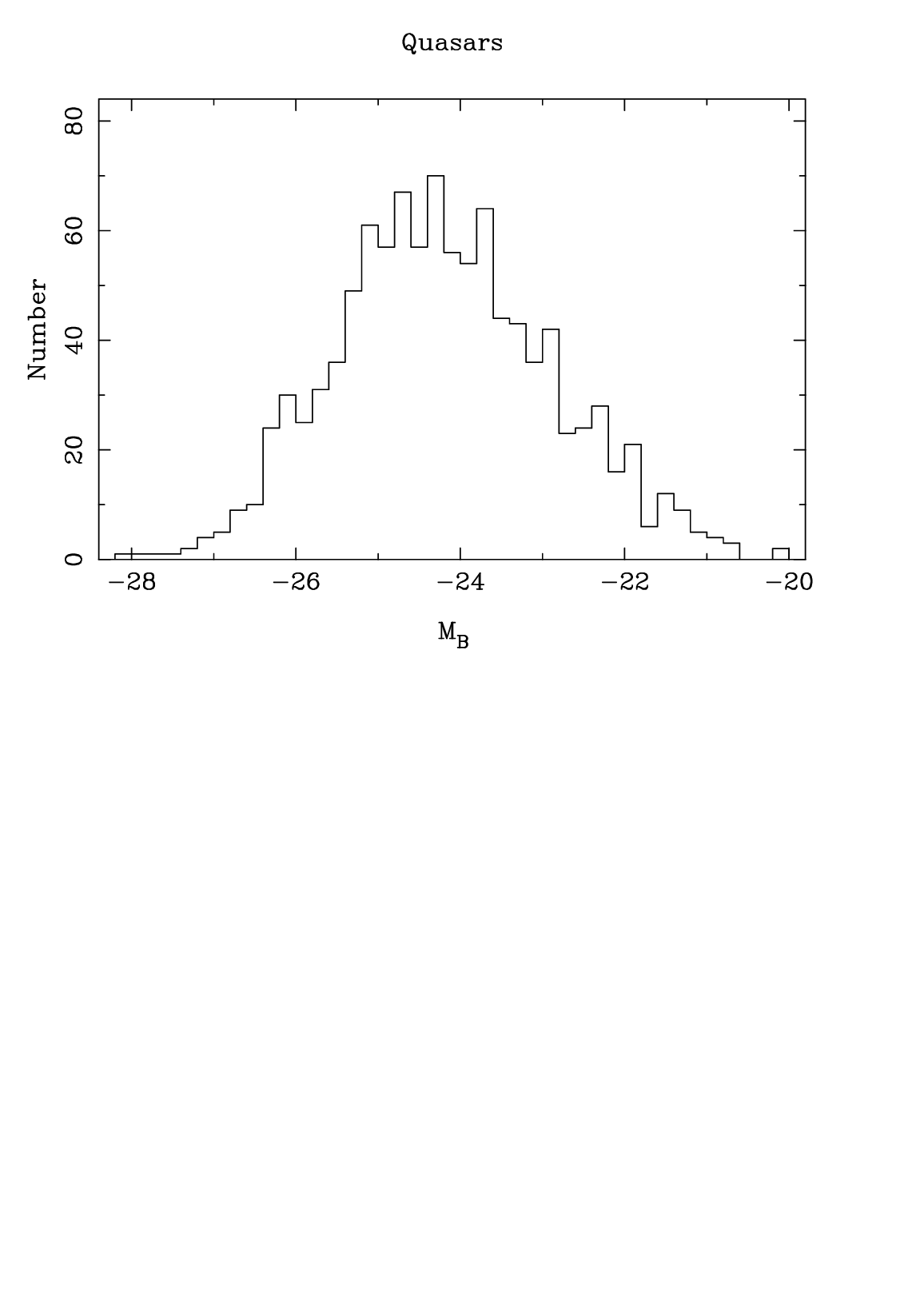}
\end{picture}
\caption{Histograms of the redshift (left hand panel) and absolute
 magnitude (right hand panel) of quasars in the monitoring survey.
 \label{fig1}}
\end{figure*}

\section{Introduction}

When quasars were first discovered in the 1960s a fundamental property
which defined their nature was the substantial variation in brightness
in optical passbands, observed over a timescale of a few months.  This
put severe constraints on the maximum size of the emitting region, and
was largely responsible for the eventual understanding of the overall
structure of an active galactic nucleus (AGN).  Since these early
observations, much effort has been put into establishing the minimum
timescale on which AGN can be seen to vary in brightness, both in
optical and X-ray passbands (\cite{kg04,st04}).  More problematic has
been the attempt to determine the longest timescale on which AGN show
significant variations.  Over the last 20 years or so a number of
groups have published the results of monitoring programmes covering a
decade or more (\cite{h02,p99,c96,h94,t94}).  Most of these programmes
have been aimed at establishing correlations between a measure of the
degree of variability, typically amplitude, and parameters such as
luminosity or redshift.  Although there is some divergence in the
results, there now seems reasonable agreement that there is an
anti-correlation between luminosity and amplitude in the sense that
more luminous quasars vary with smaller amplitudes, but there is no
convincing detection of a correlation between amplitude and redshift.
These studies do however suffer from several drawbacks, including the
lack of a satisfactory definition of amplitude which is independent of
the timespan of the observations, and the degeneracy between redshift
and luminosity which are correlated in most quasar samples.

A new approach to the analysis of AGN variability has recently become
possible with the publication of a very large quasar sample from the
Sloan Digital Sky Survey (SDSS) (\cite{sc03}).  Several groups have
used this data to analyse quasar variability, but in particular two
groups (\cite{s06,d05}) have combined this data with measures of early
epoch material from the Palomar Sky Surveys to give up to three well
separated epochs.  They have then taken advantage of an assumed time
dilation effect to recover rest frame light curves which result in a
wide range of time differences, and enable them to construct a well
sampled structure function.  This neat idea seems to work moderately
well to give the rough shape of the underlying structure function, but
there are many systematic effects which limit the amount of information
which can be extracted.  These include the challenge of correcting for
the different passbands in which the three surveys were taken, large
and hard to quantify Malmquist biases and the differing effect of the
underlying host galaxy in the different media of the three surveys.

An important parameter to be obtained from a monitoring programme is
the characteristic timescale of variation.  In those surveys where
a structure function or Fourier power spectrum has been constructed
from the light curves it has been generally found that there is a power
law rise towards longer timescales, but the question of where this
power law relation turns over has not so far been established in a
convincing way.  This break point which gives the characteristic
timescale of variation is of great importance in establishing the
overall size of the accretion disc, and can also constrain the mass
transfer rate.  In the event that variations are caused by microlensing
(\cite{h96}), it is a measure of the mass of the microlensing bodies.
The early values for timescale of variation from this monitoring
programme (\cite{h96}) were based firstly on visual estimations of a
representative value from examination of the light curves.  A second
approach involved an attempt to model the observed autocorrelation
function with simulated light curves from the literature.  Neither of
these approaches can be considered as anything more than lower limits,
especially in view of the relatively short run of data available at
that time.  The problem of timescale measurement was revisited some
years later (\cite{h01}) when Fourier power spectrum analysis was
used, along the lines of the present paper.  In this case an unbroken
power law was observed, implying that no characteristic timescale was
measureable with the available data.

A considerable amount of work has already been undertaken to measure
cut-off frequencies for AGN variation in X-ray bands (\cite{u02,e99}). 
There now seems little doubt that characteristic timescales have been
measured, which are typically about one month, with an index for the
power law of around -1.8, and the position of the break is interpreted
as a measure of the black hole mass (\cite{u02}).

In this paper we assemble the best available optical data from
timescales of a few days to thirty years.  We then investigate methods
for analysing the data with a view to measuring timescales of variation
both in the observer's frame and quasar rest frame.  From the combined
data we then construct Fourier power spectra spanning three orders of
magnitude in time with a view to measuring timescale of variation.
Finally, we review the implications of these results in the context of
various models for the observed optical variations.

\section{The quasar monitoring programme}

\subsection{Photometric observations}

This investigation into the timescale of variability of AGN is
primarily based on a 28 year monitoring programme with photographic
plates taken with the UK 1.2m Schmidt telescope at Siding Spring
Observatory in Australia.  The survey area comprises the central
unvignetted part of ESO/SERC field 287 centred on 21h 28m,
-45$^{\circ}$ (1950) and covers approximately 20 square degrees.  Over
300 plates were taken of this field during the lifetime of the UK
Schmidt, in a variety of passbands from $U$ to $I$, as well as
objective prism and other more specialised plates, but for the purposes
of the present project we restrict our attention to a series of plates
taken every year for 26 years in the $B_{J}$ (IIIa-J/GG395) passband
from 1977 to 2002, and for 23 years in the $R$ (IIIa-F/RG630) passband
from 1980 to 2002.  In most years 4 plates were obtained, and details
of all but the last year are given by \cite{h03}.

The plates were measured by the COSMOS and SuperCOSMOS machines at the
Royal Observatory Edinburgh, which produced a catalogue of
approximately 200,000 objects to a completeness limit of
$B_{J} = 21.5$.  There was also a period of 15 years from 1983 to 1997
where there was a homogeneous run of data comprising 4 good $B_{J}$
plates every year, and these were digitally stacked to provide a second
dataset with a completeness of about $B_{J} = 22$.  The plates were
calibrated using deep CCD sequences ({\cite{h96}), and reduced to the
same zero-point using local photometric transformations.  The
photometric accuracy on each measurement was obtained from repeat
measures of the same objects on several different plates, and found to
be $\pm 0.15$ magnitudes in the magnitude range appropriate to the
quasar sample.  The overall rms variation on the light curves,
including mean magnitudes where more than one plate per year was
available was $\pm 0.11$ magnitudes.  This was evaluated directly by
measuring the root mean square (rms) variation of the light curves of
samples of stars of similar colour and apparent magnitude distribution,
assumed to be non-variable.

\subsection{Sample selection}

The sample of candidate quasars or AGN for spectroscopic confirmation
were selected by a number of methods.  These included ultraviolet
excess, typically with a limit of $B_{J}-U < 0.0$, and variability
with an amplitude over 28 years of $\delta B_{J} > 0.35$.  The exact
criterion for variability is given in \cite{h96}.  In addition to these
basic selection criteria, other methods were used including objective
prism searches, and candidates from radio surveys with the Molongolo
and Australia telescopes, but in practice all such candidates were
already selected by other methods.  Variability has proved to be a
particularly powerful way of detecting quasars, and it has been shown
(\cite{h00}) that nearly all quasars can be detected on the basis of
variability after 15 years of survey data.

Separate sampling programmes were carried out with the stacked and
single plate data.  Although there was an approximate 75\% overlap
between the two selection lists, the stacked data contained fainter
objects below the limit of the single plate data, whereas the single
plate data had less of an overcrowding problem and contained objects
which were merged in the stacked data.  In all there were about 2000
good candidates suitable for spectroscopic observations to establish
whether they were AGN, and if so to measure their redshifts.

\subsection{Spectroscopy}

Spectroscopic follow up of candidates from the monitoring programme was
commenced some 20 years ago, and as the wide field capability of
spectrographs has improved the number of confirmed AGN has rapidly
increased.  The first sample to be analysed  (\cite{h96})
comprised some 400 quasars with redshifts.  This had increased to
around 600 for the analysis by \cite{h02}, as a result of a number of
small time allocations with single slit spectrographs.
The most recent and largest award of time was in July
2002 with the 2dF multi-object spectrograph on the Anglo-Australian
Telescope.  This resulted in a homogeneous set of redshifts for about
1000 AGN, including a number of re-observations, and brought the total
number of confirmed AGN with redshifts to about 1500 in the 20 square
degrees of the survey.  The redshifts were measured to an accuracy of
approximately $\pm 0.2\%$, evaluated from multiple observations of the
same object; more details of the procedures used are given by 
Hawkins (2004).

The above procedure resulted in a sample of 814 quasars for which light
curves in the $B_{J}$ passband covering 26 yearly epochs were measured.
This included the requirement that there should be no missing epochs.
The redshifts of the quasars in the sample range from 0.15 to 3.8, and
the luminosities from $M_{B} = -20$ to $M_{B} = -28$.  The
distributions of these two quantities are shown in the form of
histograms in Fig.~\ref{fig1}.  Light curves in $R$ were also available
for the quasar sample, but the restriction that there should be no
missing epochs, combined with the blue colour of most quasars, meant
that the sample of acceptable red passband light curves was reduced
to 403.  These two sets of light curves formed the basis for the
timescale analysis described below.

\begin{figure*}[t]
\setlength{\unitlength}{1mm}
\begin{picture} (200,45) (0,45) 
\includegraphics[width=0.5\textwidth]{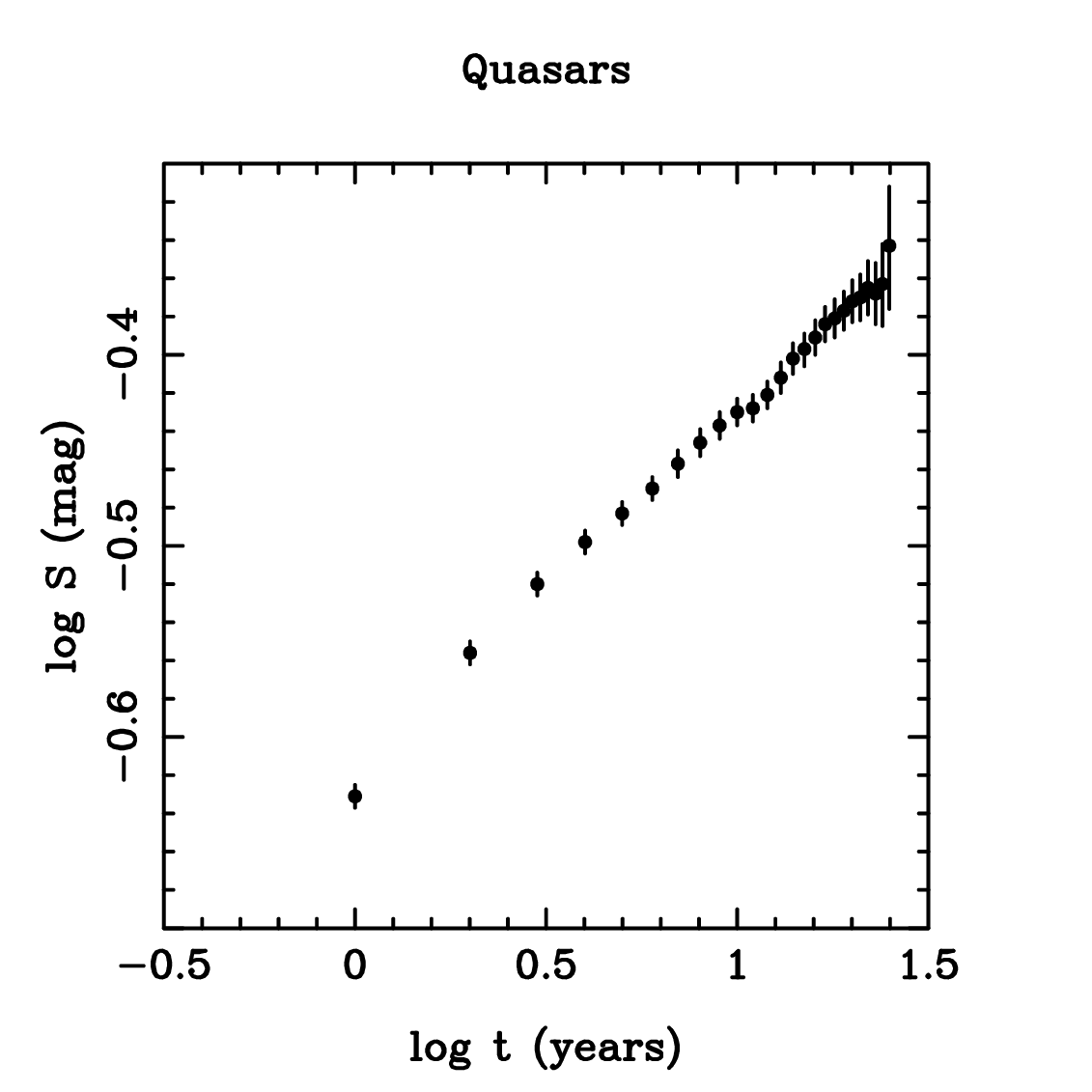}
\end{picture}
\begin{picture} (200,45) (-90,0) 
\includegraphics[width=0.5\textwidth]{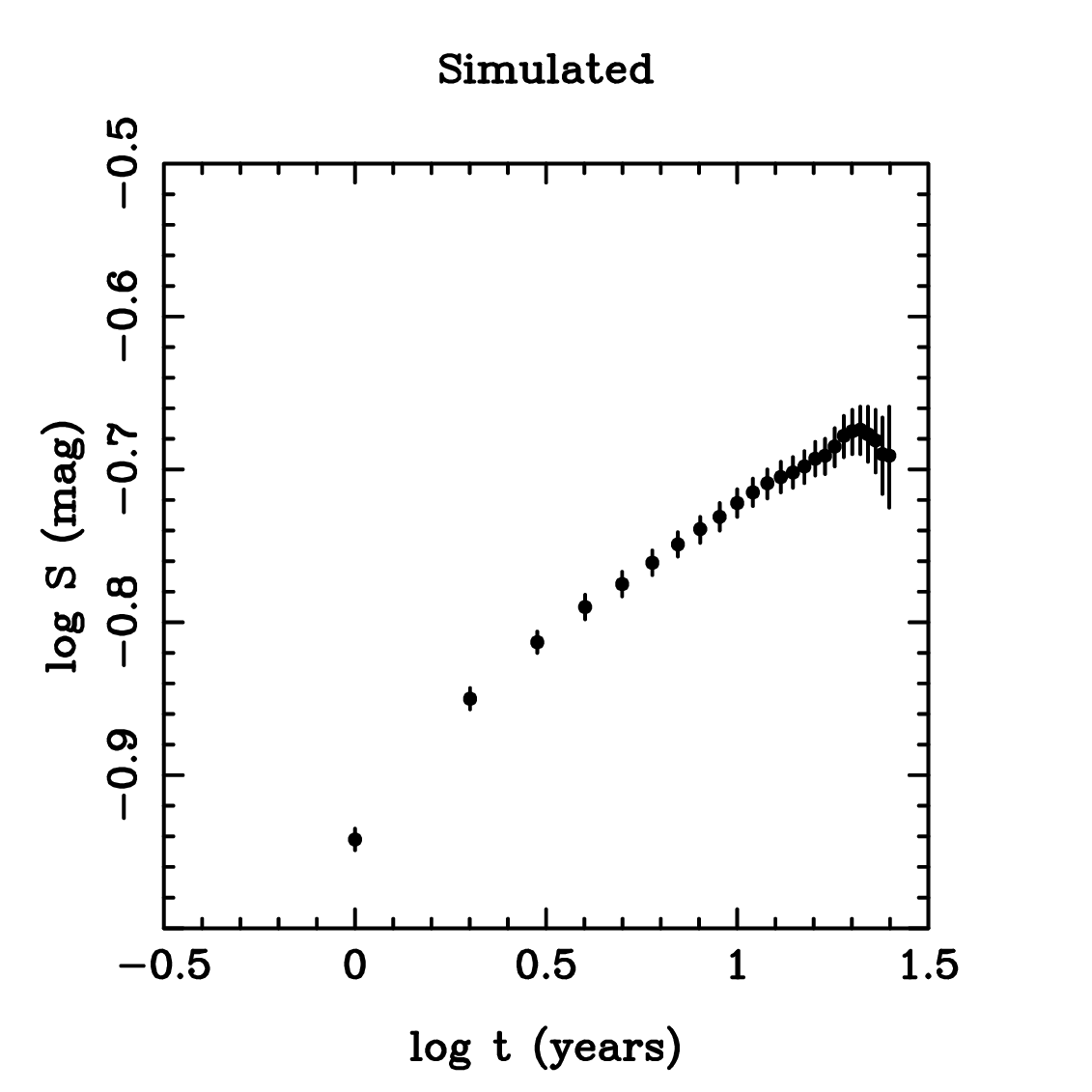}
\end{picture}
\begin{picture} (200,45) (0,45) 
\includegraphics[width=0.5\textwidth]{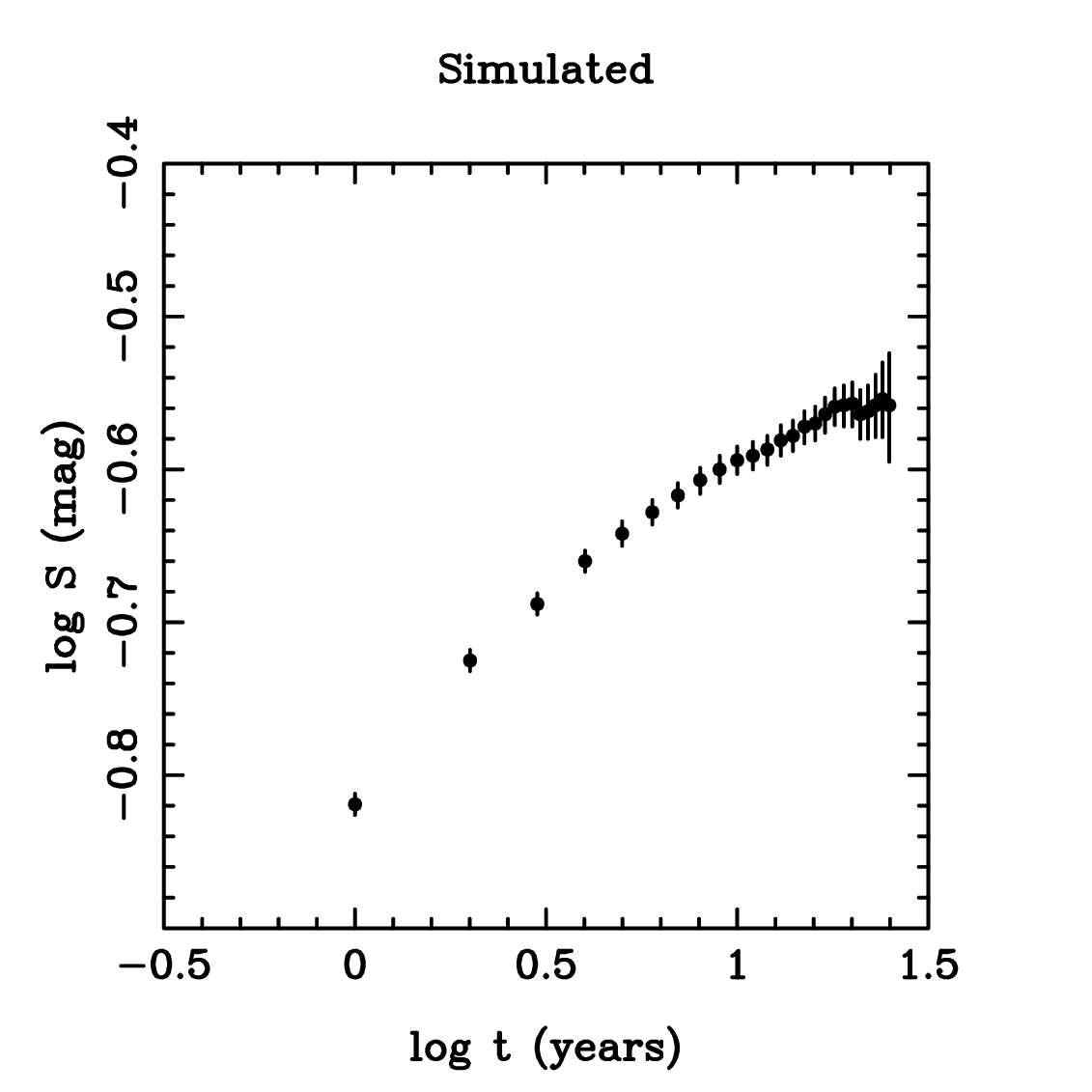}
\end{picture}
\begin{picture} (200,45) (-90,0) 
\includegraphics[width=0.5\textwidth]{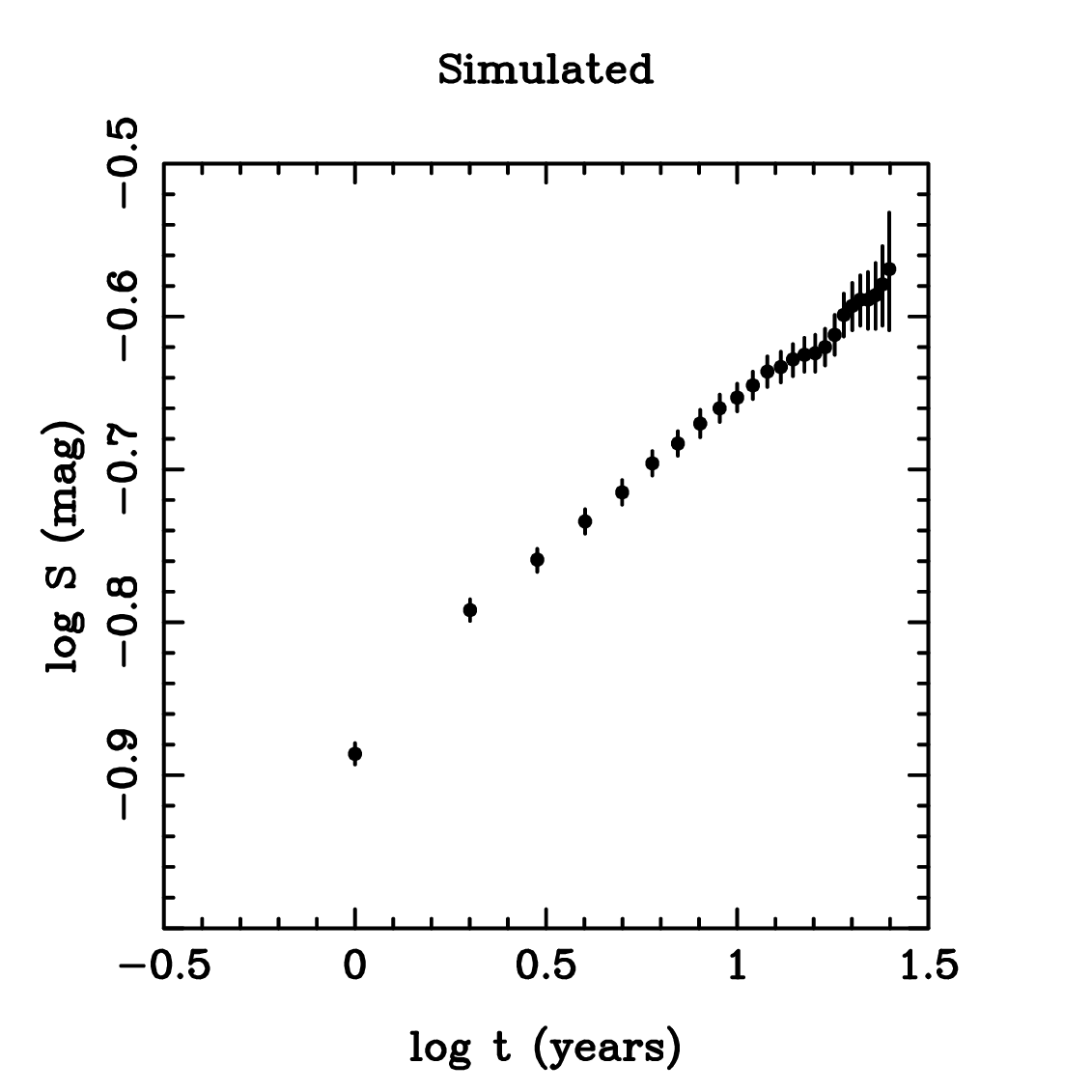}
\end{picture}
\caption{Structure functions for the sample of 814 quasars in
 Field 287 (top left hand panel) and for three realisations of 814
 simulated light curves with a similar power spectrum of variations
 and the same sampling.
 \label{fig2}}
\end{figure*}

\subsection{The MACHO quasars}

In earlier work on the sample of quasars in field 287 described above,
a major limitation to the measurement of timescales has been the length
of the baseline in the log/log plot (e.g. \cite{h01,h02}).  To date
the approach has been to extend as far as possible the length of the
time baseline, but the logarithmic nature of the increase makes this a
strategy of diminishing returns.  One way of improving the chances of
detecting a turnover is to provide a longer baseline by including
observations taken over shorter timescales.  An ideal database for this
purpose is the long set of observations taken for the MACHO project.
The idea of this well-known project was to monitor several million
stars in the direction of the Magellanic Clouds with a typical
frequency of a few days to look for the characteristic signature of a
Galactic microlensing event (\cite{a97}).

Although the overwhelming majority of the objects in the MACHO
monitoring programme were stars, there were inevitably a number of
quasars in the area of interest which were also observed.  These have
recently been collated and published with their redshifts (\cite{g03}),
and form a very useful, if rather small, database for measuring short
timescale variations down to a few days.  There is a total of 59
quasars in their database, including their new discoveries and those
already known.  The redshifts and absolute magnitudes of these quasars
lie in the ranges $0.1 < z < 2.8$ and $-28 < M_{B} < -20$, and follow
similar distributions to those shown in Fig.~\ref{fig1}.

The observations were made with a CCD camera in $B$ and $R$ bands on
every night permitted by the weather and other logistical constraints.
Occasionaly more than one observation per night was obtained.  The
photometric accuracy is recorded for each observation, and is quite
variable depending on seeing conditions, level of contamination by
neighbouring objects and the brightness of the object.  A typical
error estimate on a single measurement is $\pm 0.1$ magnitudes.  For
the purposes of the work in this paper, the observations were put into
bins of 10 or 50 days, and the errors combined in quadrature.  For the
50 day bins this resulted in a typical error per epoch of about
$\pm 0.03$ magnitudes.

Although the full set of MACHO observations covers more than 7 years,
the patchiness of the coverage for some quasars combined with the
requirement for uninterrupted sampling meant that for the 50 day bins
the original total of 59 quasars was reduced in the blue passband to
48, with a continuous run of 51 epochs, and in the red to 35 over 36
epochs.  For the 10 day bins useful data was only obtained in the
periods of relatively intense observations due to the requirement that
there should be a continuous run of data.  This meant that the total
was reduced to 32 quasars in the blue passband and 25 in the red, both
measured over 33 epochs.  These were the light curves used for the
analysis described below.

\section{Time series analysis}

It has been known for some time (\cite{c96,h96}) that the spectrum of
variations of AGN, when measured over several years, increases in
power towards longer timescales.  The Fourier power spectrum of the
variations has been shown to be close to a power law in form
(\cite{h01}), but the question of where this power law turns over at
long timescales, and hence the identification of a characteristic
timescale, has remained open.  For example, Hawkins (2001)
concluded that the Fourier power spectra of quasar light curves were
adequately fitted by a power law relationship with no turnover,
implying that the run of data was still too short to reach any
characteristic timescale.  In another paper (\cite{h02}), structure
functions were used to match the presentation of simulated data, but
due to windowing effects producing apparently spurious features it was
only possible to compare slopes, and not any more detailed features.

It is clear that the measurement of timescales in light curves requires
a statistical treatment of the data which enables the detection of a
break or other feature in the spectrum of variations, combined with a
sufficiently long run of photometric measurements.  To date, two
functional forms have principally been used for the analysis of
timescales, each of which has points in favour and against.  We discuss
and illustrate these in the following two subsections.

\subsection{The structure function}

The most frequently used functional form for the analysis of quasar
variations is undoubtedly the structure function
(\cite{v04,d03,h02,c96,t94}), together with the closely related
auto-correlation function (\cite{h96}).  In addition, theoretical
models of quasar variations have been published in the form of
structure functions (\cite{k98,h02}).  The structure function $S(\tau)$
is usually defined as

\begin{equation}\
S(\tau) =\sqrt{\frac{1}{N(\tau)}\sum_{i<j}[m(t_{j})-m(t_{i})]^{2}}
\end{equation}

\noindent
where $m(t_{i})$ is the magnitude measure at epoch $t_{i}$, and the
sum runs over the $N(\tau)$ epochs for which $t_{j}-t_{i} = \tau$, and
only includes those time differences where data is available.  If the
structure function is for more than one object, then the summation runs
over time intervals for all the light curves in the sample.  This
results in one of the most important advantages of the structure
function over other methods, that is its insensitivity to gaps in the
run of data, especially on short timescales.  This has proved of great
benefit to surveys which rely on archival observations taken at
irregular intervals for their data.

To set against their usefulness for analysing unevenly sampled data
sets, there are some serious problems associated with structure
functions.  Perhaps the best known is the difficulty of error analysis,
associated with the correlation of errors from point to point.  This
has the effect of producing what at first sight appear to be real
features in the form of changes of slope or breaks, and makes the
detailed interpretation of the shape of a structure function very
difficult.  To illustrate the problem, we show in the top left hand
panel of Fig.~\ref{fig2} the structure function for the sample of
quasars in Field 287.  The plot is constructed by calculating the mean
structure function for the sample of 814 blue passband light curves, as
described above.  In the log/log plot, the structure function increases
to longer timescales in a roughly linear mannner, indicating a power
law relationship.  There is no sign of a significant turn over or
break, although there is the possibility of features associated with
the wavy departures from a strictly linear increase.

In order to investigate the significance of the shape of the quasar
structure function, simulated light curves were constructed with known
statistics of variation.  Each simulation was made by defining a pure
power law power spectrum for a large number of data points, taking the
square root, applying a random phase and taking the inverse Fourier
transform.  The index of the power law was -1.2, chosen to give a
structure function slope close to that measured for the quasars.  This
produced a long light curve with the required variability
characteristics.  Out of this were cut 814 individual light curves of
the same length (26 epochs) as for the observations described in
section 2.  Structure functions were then calculated for the sets of
light curves as described above, three of which are shown as log/log
plots in the remaining panels of Fig.~\ref{fig2}.

It will be seen that for short timescales the simulated structure
functions increase with time in a roughly linear fashion in all three
cases.  However, beyond about 5 years the three functions diverge, with
one eventually turning over at longer timescales while the other two
continue to rise at different rates, with some fluctuations.  If the
only structure function available were the one in the top right hand
panel, a naive interpretation would suggest a power law relationship
with a long timescale cut-off.  This interpretation is made more
plausible by predictions in the literature for such behaviour in AGN
structure functions for a wide range of models (\cite{h02,k98}).  There
is also the possibility of other features associated with the wavy
departures from a strictly linear increase.  In fact, it is clear that
from the same power law spectrum, a variety of structure function
shapes can result. The data in the top left panel of
Fig.~\ref{fig2} is largely the same as that used in Hawkins (2002),
and has a very similar shape.  It illuminates the point made in that
paper that the only reason for working with structure functions is that
the simulations were published in that form.

It is easy to show that for an infinite run of observations, the
structure function for a power law spectrum of variations is also a
power law.  The simulated structure functions in Fig.~\ref{fig2} 
clearly depart from power law shapes, especially towards long
timescales.  This is due to the unpredictable effect of windowing on a
finite run of data, which produces different morphologies for different
sets of light curves, despite their being drawn from the same
underlying spectrum of variations.

The data in Fig.~\ref{fig2} make clear the unsuitability of structure
functions for the analysis of short runs of data, and especially for
the identification of cut-offs to be used for the measurement of
timescale.  It is of course true that because of irregular sampling in
many sets of data, it is the only feasible choice.  In this case it is
probably best to confine the analysis to an average measurement of the
power law slope which is less affected by windowing effects, although
the presence of noise in the light curves will alter the observed
index.

\begin{figure*}
\setlength{\unitlength}{1mm}
\begin{picture} (200,90) (-10,80)
\includegraphics[width=0.9\textwidth]{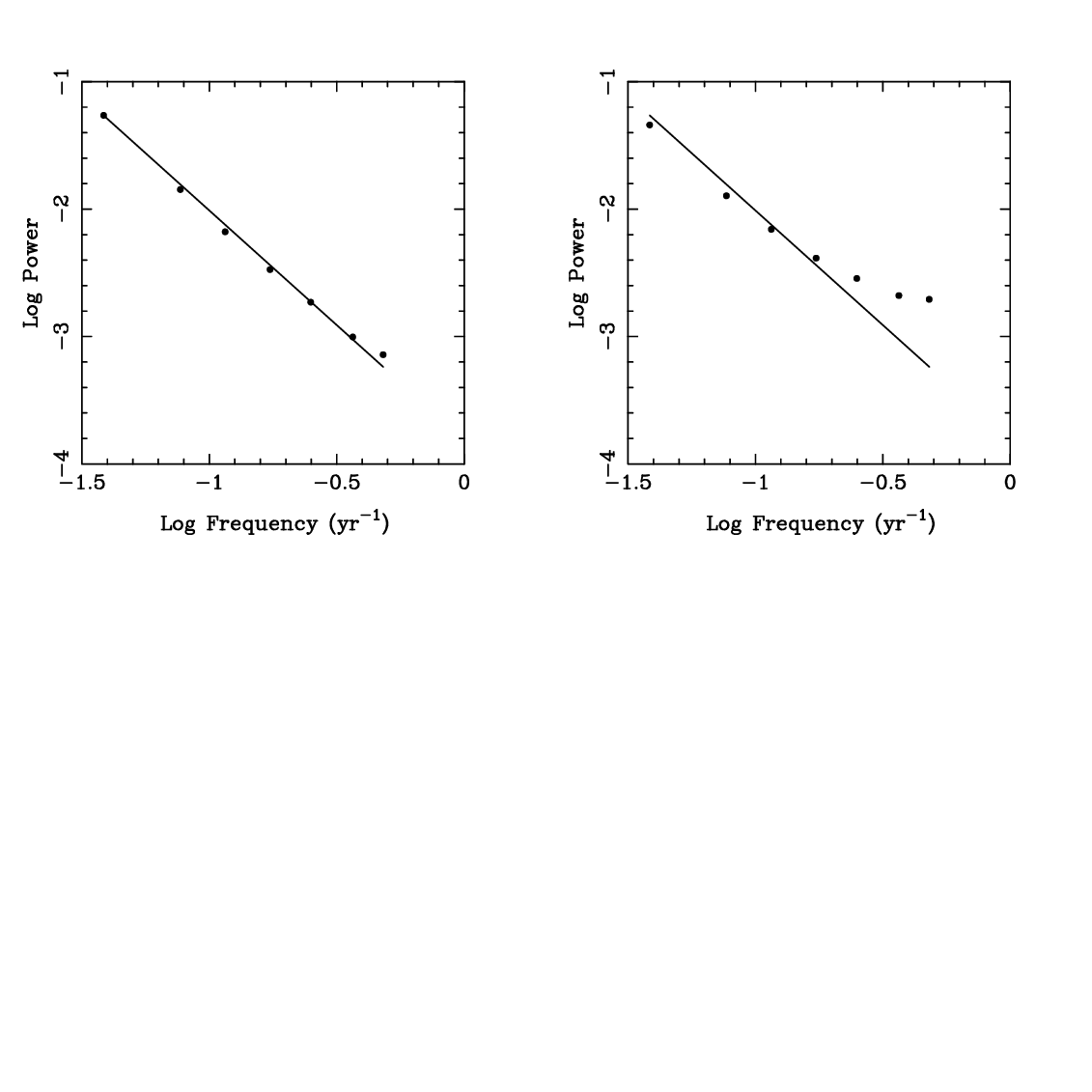}
\end{picture}
\caption{Discrete Fourier power spectra for simulated light curves
 constructed from a power law power spectrum of index -1.8, indicated
 by the solid line (left hand panel).  The effect of adding Gaussian
 noise with standard deviation 0.06 magnitudes is shown in the right
 hand panel.
 \label{fig3}}
\end{figure*}

\subsection{The Fourier power spectrum}

For an infinite run of data the Fourier power spectrum and structure
function or auto-correlation function are essentially equivalent,
one being the Fourier transform of the other.  However, for finite
datasets, and especially for the short datasets typically encountered
in AGN monitoring programmes, the windowing function has a significant
and very different effect in the two cases.  We define the Fourier
power spectrum P(s) as

\begin{equation}
P(s_{i}) = \frac{\tau}{N} \left( \sum_{j=1,N} m(t_{j}) cos \frac{2 \pi
 i j} {N}\right)^{2} + \frac{\tau}{N} \left( \sum_{j=1,N}
 m(t_{j}) sin \frac{2 \pi i j} {N}\right)^{2}
\end{equation}

\noindent
where i runs over the N equally spaced epochs of observation separated
by time $\tau$, and $m(t_{j})$ is the magnitude at epoch $t_{j}$.  In
the case of a sample of light curves, the integration for each
frequency continues over all sample members.

The main advantage of the structure function, its insensitivity to
irregular sampling, is not shared by Fourier transforms.  Here,
unevenly sampled data leads to aliassing in the form of spurious
features which can cause great difficulties in the measurement and
interpretation of power spectra.  However, given evenly spaced
observations, errors are straightforward to calculate as each Fourier
frequency is essentially independent, and the shape and features of
a power spectrum are relatively easy to interpret.  The windowing
function due to short runs of data produces little distortion to the
power spectrum, and Gaussian noise adds a constant term to the power
spectrum, the size of which can be derived from the photometric errors
and subtracted as a correction.

The effects described in the previous paragraph are illustrated in
Fig.~\ref{fig3} which shows a Fourier power spectrum for simulated
light curves constructed as described in the previous sub-section.
The solid line indicates the power law power spectrum used for the
simulations, in this case with an index of -1.8.  The power spectrum of
the 814 light curves of 26 epochs are shown as filled circles in the
left hand panel of Fig.~\ref{fig3}.  It will be seen that very little
deviation from the original power spectrum has been produced by the
windowing process.  The right hand panel shows the effect of adding
Gaussian noise with a standard deviation of 0.06 magnitudes to the
light curves.  The effect on the power spectrum is to make a relative
increase in the power of the short timescale frequencies, which can be
corrected for by subtracting a constant from all frequencies, as
described in the previous paragraph.

\begin{figure*}
\setlength{\unitlength}{1mm}
\begin{picture} (200,90) (-10,80)
\includegraphics[width=0.9\textwidth]{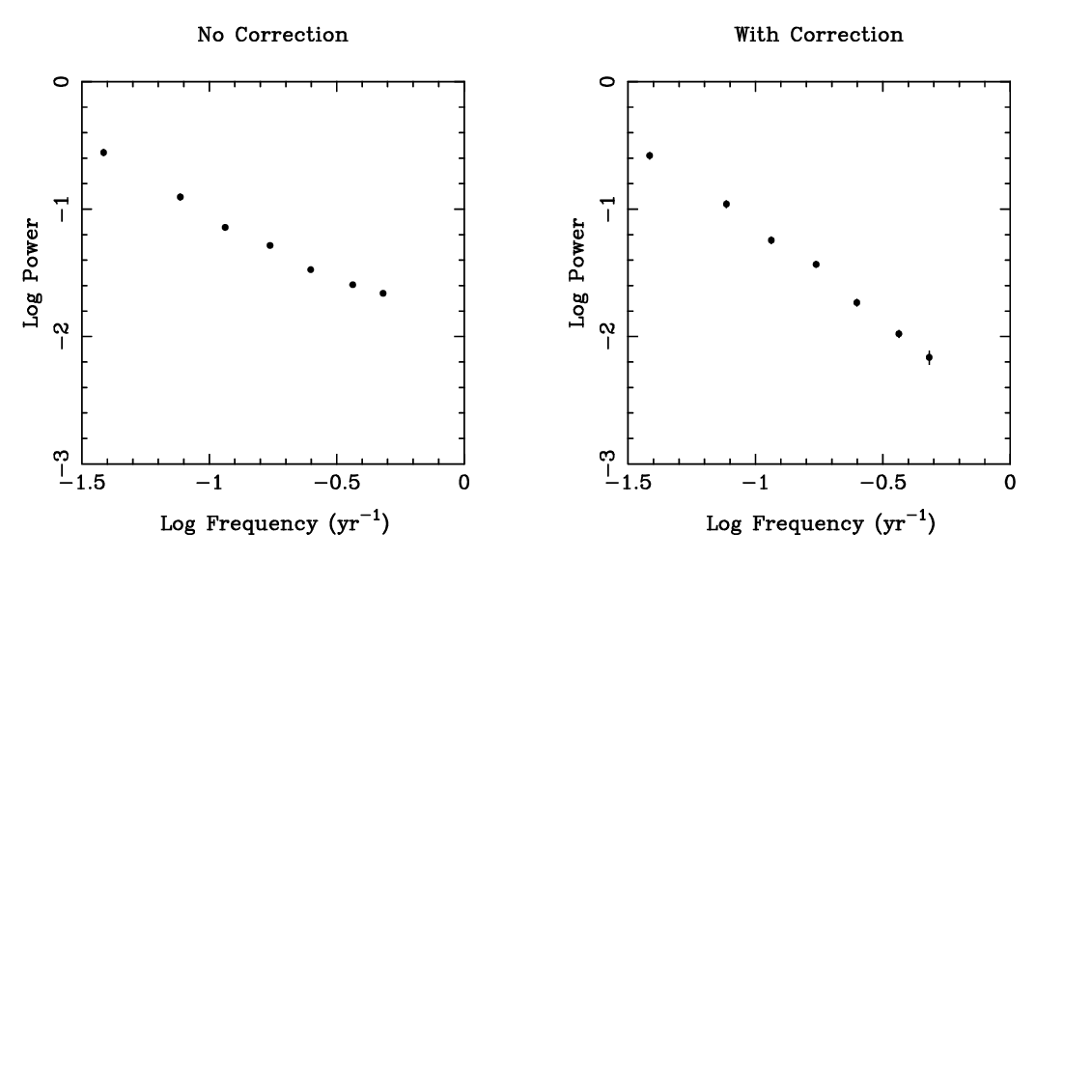}
\end{picture}
\caption{Fourier power spectra for the sample of 814 quasars in
 Field 287 for which the structure function is shown in
 Fig.~\ref{fig1}.  The left hand panel shows the power spectrum as
 measured, and the right hand panel after correcting for measurement
 noise, as described in the text.
 \label{fig4}}
\end{figure*}

\section{The timescale of quasar variation}

In the previous section we have discussed the relative merits of
structure function and Fourier power spectrum analysis, and there seems
little doubt that for the purpose of analysing the data presented here
to measure timescales, the latter approach is preferable.  Accordingly,
for the remainder of this paper we shall confine attention to Fourier
power spectra of the data.

In the measurement of timescales in the AGN light curves, we shall
consider two situations.  Firstly we shall analyse Fourier power
spectra for light curves as measured, that is in the observer's frame
where no correction is made for effects of time dilation.
Astrophysically, this corresponds to a situation where the variations
do not originate in the quasar itself, but along the line of sight at
much lower redshift.  In particular, if the variations are caused by
the microlensing effect of a cosmologically distributed population of
compact bodies (\cite{h93,h96}), the most probable redshift for lensing
peaks strongly at $z \approx 0.5$ (\cite{t84}), with only weak
dependence on the redshift of the source.  Thus the effects of time
dilation will be small, and the observer's rest frame will be the
most appropriate for measuring timescales.  Secondly we shall carry out
the analysis in the rest frame of the AGN, which is appropriate for
intrinsic variations, and will necessitate a correction to remove the
effects of time dilation to be applied to each light curve before the
timescale is measured.

In fact, AGN are known to vary at least at some level both
intrinsically and due to microlensing.  Studies of low redshift
quasars where there is no question of microlensing show variations in
brightness at least at some level (\cite{h04}), while gravitationally
lensed systems such as the double quasar Q0957+561 show unambiguous
evidence for microlensing (\cite{p98}).

\subsection{Timescale in the observer's frame}

Fig.~\ref{fig4} (left hand panel) shows the Fourier power spectrum for
blue passband light curves for the sample of 814 quasars in Field 287,
described above.  For each quasar 26 epochs of observation are
available, as described in section 2.  The Fourier power spectrum for
the sample of light curves was evaluated as described above for the
simulated data in the previous section.  The error bars are derived
from the dispersion of the individual power spectra. The shape of the
power spectrum appears close to that of a power law, that is linear in
the log/log plot, but with a noticeable decrease in slope towards high
frequencies.

In order to provide an accurate measurement of the true shape of the
power spectrum, it was necessary to correct for the effects of
measurement noise.  This was done as described at the end of the
previous section based on the measurement errors derived in section 2.
As a further check, the correction was obtained by measuring the power
spectrum of a sample of stars assumed to be non-variable, thus
providing a direct empirical measure of the measurement noise.  These
two approaches to error estimation gave similar results, and the right
hand panel of Fig.~\ref{fig4} shows the result of making this
correction, which basically has the effect of straightening and
steepening the linear form of the power spectrum.

\begin{figure*}[t]
\setlength{\unitlength}{1mm}
\begin{picture} (200,170) (-10,0)
\includegraphics[width=0.9\textwidth]{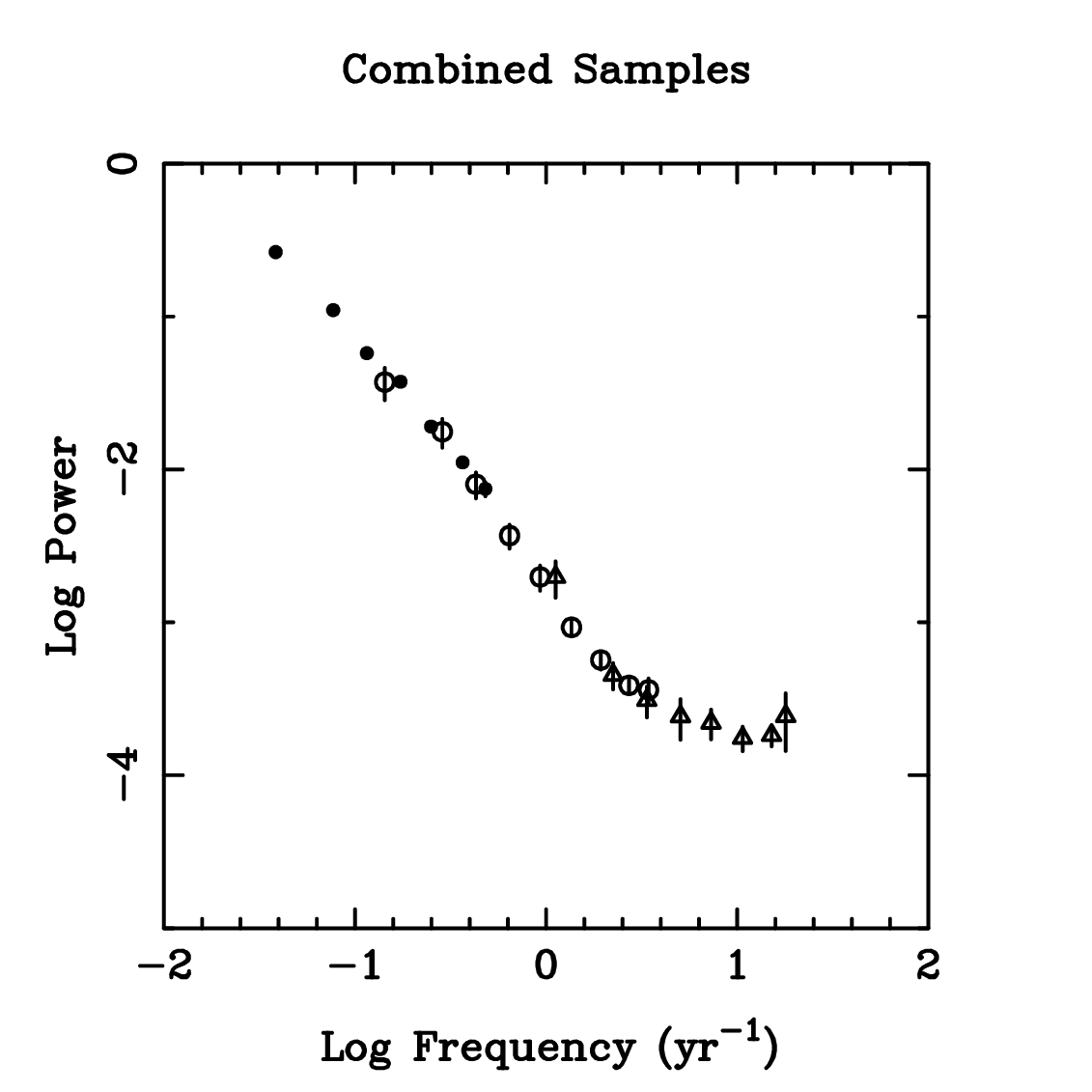}
\end{picture}
\caption{Fourier power spectra of blue passband light curves in the
 observers frame for the sample of quasars in Field 287 sampled yearly
 (filled circles), the  MACHO quasars with 50 day sampling (open
 circles) and 10 day sampling (open triangles).
 \label{fig5}}
\end{figure*}

\begin{figure*}[t]
\setlength{\unitlength}{1mm}
\begin{picture} (200,45) (0,45)
\includegraphics[width=0.5\textwidth]{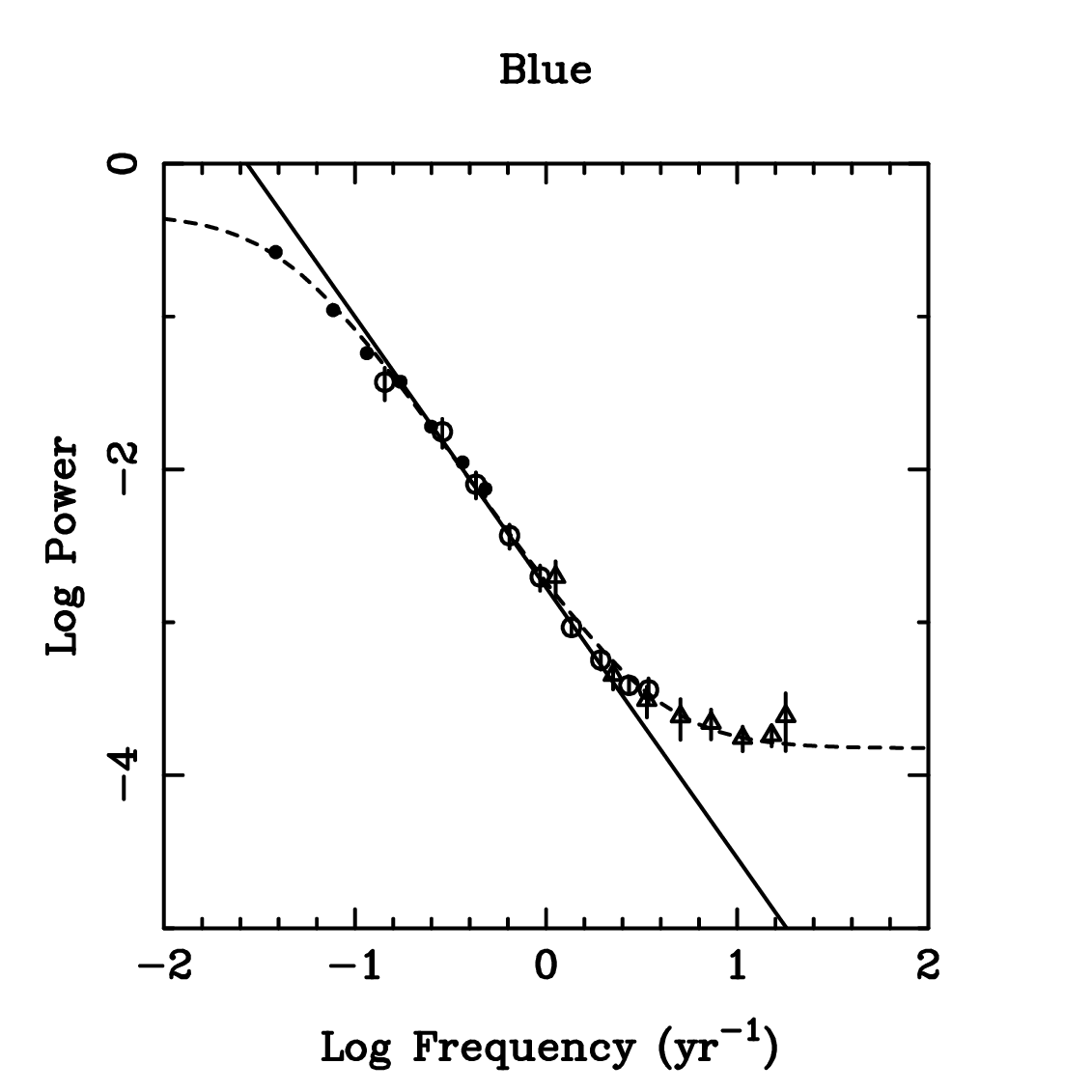}
\end{picture}
\begin{picture} (200,45) (-90,0)
\includegraphics[width=0.5\textwidth]{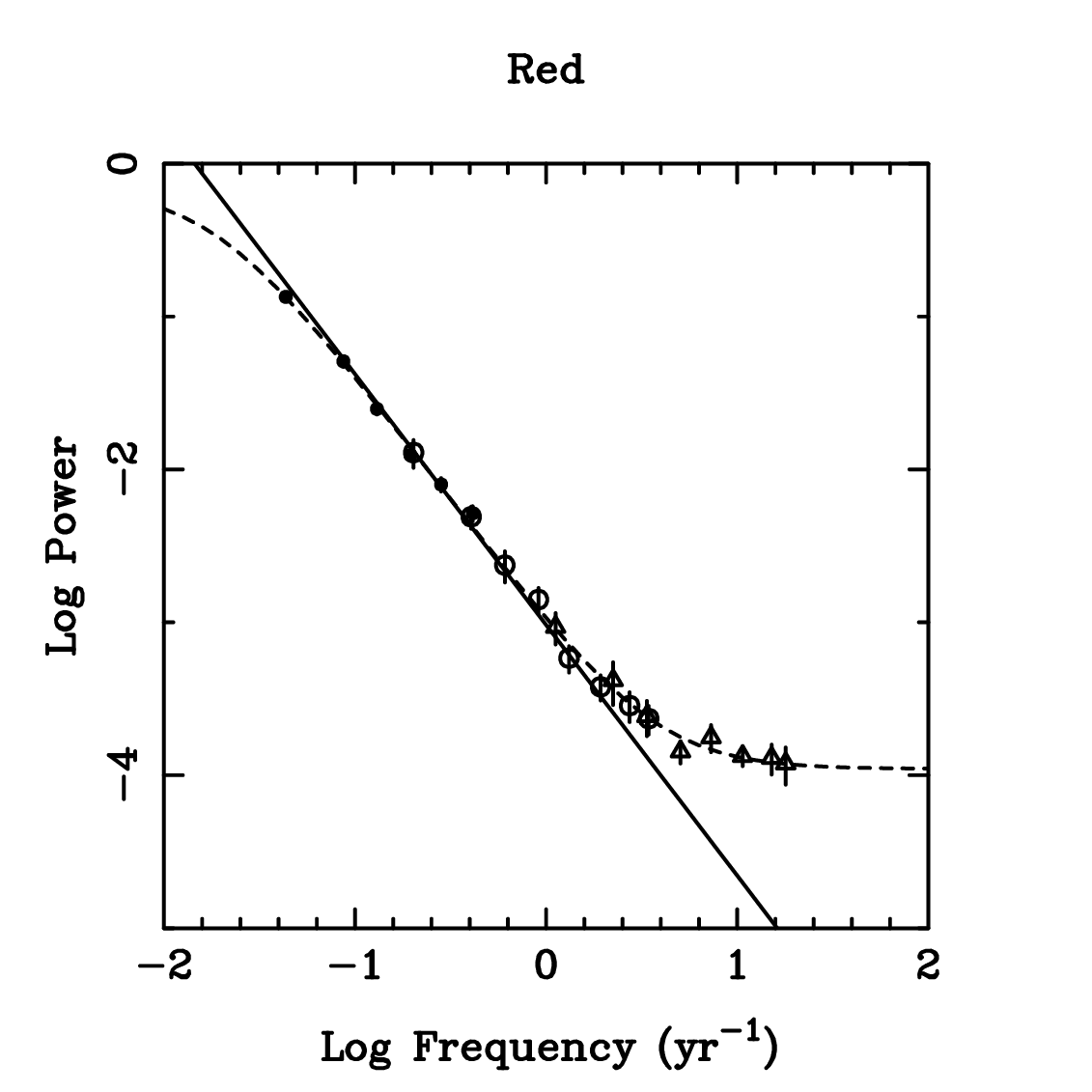}
\end{picture}
\caption{Fourier power spectra of blue passband light curves (left hand
 panel) and red passband light curves (right hand panel) in the
 observer's frame for quasars in the Field 287 and MACHO samples.
 Symbols are as for Fig.~\ref{fig5}.  The dashed line is the best-fit
 curve of Eq.\ (3), and the solid line is the asymptotic power law
 relation for $P(f)$ as described in the
 text.
 \label{fig6}}
\end{figure*}

The power law increase in the power spectrum in Fig.~\ref{fig4} implies
no preferred or characteristic timescale for the quasar variations.
Detailed examination of the long timescale end does in fact suggest
the possibility of a break in the slope, but given the relatively
short baseline in frequency it is not possible to establish its
significance in a formal way.  The monitoring programme has already
been running for nearly thirty years, and given the logarithmic nature
of a power law relation, the amount of extra time needed to make a
significant improvement in the frequency baseline is comparable to a
human lifetime.  An alternative approach is to add in shorter timescale
measurements to provide a firmer measurement of the asymptotic slope
in the log/log plot towards high frequencies.  The MACHO observations
described in section 2 provide an excellent if somewhat small database
for this purpose.

Fig.~\ref{fig5} shows Fourier power spectra for the combined Field 287
and MACHO data.  The data from the right hand panel of Fig.~\ref{fig4}
are shown as filled circles.  The open circles show the power spectrum
for the MACHO data with 50 day sampling interval, constructed as for
the Field 287 sample.  The correction for the windowing effect obtained
from the errors in the MACHO data was applied to the to the 50-day
sampling power spectrum as discussed above, although it was found to be
small due to the relatively small errors in the CCD data.  The
situation was rather different for the 10-day sampled data.  Here for
the shorter timescales the random errors were comparable in size to the
signal, and the correction could not be made with any confidence.  The
data was therefore left unchanged and modelled in the fitting procedure
described below.  There was no arbitrary
normalisation applied to any of the power spectra, and the fact that
in the area of overlap they are coincident is solely attributable to
the homogeneity of the data and reduction procedures. The third set of
data in Fig.~\ref{fig5}, shown as open triangles, is from the same set
of MACHO observations but with 10 day sampling.  The Fourier power
spectrum is produced in the same way as for the data with 50 day
spacing.  Inspection of Fig.~\ref{fig5} shows that all three sets of
data points consistently map out the same curve, both in the sense of
continuity of slope, and agreement in areas of overlap.  This is
perhaps not too surprising for the two different samplings of the MACHO
data, but the very good agreement between these and the long term
data from Field 287 implies that sample selection effects are not
significantly distorting the underlying spectrum of variability.

The overall shape of the function in Fig.~\ref{fig5} comprises a linear
section in the middle, which flattens to a constant at frequencies
corresponding to timescales of less than about 30 days, which appears
to be the effect of measurement noise from the 10 day sample for which
no correction was made.  Towards low frequencies the power spectrum
appears to maintain a linear (power law) relation, with only a slight
indication of a turn over.  In order to quantify a possible change of
slope, and hence an associated timescale, we use the function

\begin{equation}
P(f) = \frac{C}{\left(1+\frac{f}{f_{c}}\right)^{a}}+d
\end{equation}

\noindent
This is a small modification of a function employed for the purpose
of measuring timescales of X-ray variabilty (\cite{e99}).  The function
describes the power spectrum $P(f)$ of a power law relation with index
$a$ at high frequencies which breaks to a constant value $c$ at a
cut-off frequency $f_{c}$.  The modification is to add a constant term
$d$ to model the flattening attributed to noise at high frequencies.
The function $P(f)$ was fitted iteratively to the data by varying the
four parameters $a$, $f$, $c$ and $d$ and looking for a minimum
$\chi^{2}$, which was measured as 28.3 with 23 degrees of freedom.  The
best fit for $P(f)$ is shown as a dashed line in the left hand panel of
Fig.~\ref{fig6}.  Also shown, as a solid line, is the asymptotic power
law relation for $P(f)$.  This has index $a$, with best-fit value
$a = -1.77$.  The 'cut-off' frequency $f_{c}$, which is of central
importance to this investigation, has a best fit value corresponding in
the time domain to $24.0 \pm 2.7$ years.  Given that the total run of
data is only 25 years,  this figure should be treated as a lower limit
to the timescale of variation.  The error estimates were
obtained by selecting a large number of sub-samples of the data and
carrying out the same parameter estimation procedure as for the full
sample.  The rms dispersion in the measurement of the cut-off frequency
$f_{c}$ was then taken as the error.

In addition to the light curves in the $B_{J}$ passband, light curves
were also available in $R$. These observations, described in section 2,
are somewhat inferior to the $B_{J}$ observations in several respects.
Firstly, they were only started in 1980, and so comprise 23 epochs as
opposed to 26.  Secondly, due to the blue colour of most of the
quasars, a significant number were not included in the $R$ sample
because they were not detected in the $R$ band in one or more epochs. 
This resulted in a sample of 403 quasars with complete $R$ band light
curves compared with 814 for the $B_{J}$ band.  Another problem arising
from the blue colour of most quasars is that the $R$ band measures tend
to be closer to the the plate limit, and hence noisier.  Finally, the
smaller amplitude of many AGN in the $R$ band (\cite{h03}) means that
there is less power in the Fourier power spectrum, and hence poorer
signal-to-noise in the detection of any features.  Despite these
limitations, we have analysed the red passband light curves as for the
blue, and show the results in the right hand panel of Fig.~\ref{fig6}.
The best fit value for the cut-off frequency in the time domain is
$70.9 \pm 33.4$ years.  Given that the data only cover 22 years, we
shall take that as a lower limit for the timescale of variation.

\subsection{Timescale in the quasar rest frame}

The data in Fig.~\ref{fig6} are all in the observer's rest frame, with
no correction for any effect of time dilation, as discussed at the
beginning of this section.  However, if the observed variation is
intrinsic to the quasars at cosmological distances, one would expect a
time dilation effect, and for the effect to be large.  In order to
allow for time dilation and reduce the observations to the rest frame
of the quasar, it is necessary to change the timescale of variation of
each individual quasar of redshift $z$ by a factor $(1+z)^{-1}$, and
hence the frequency scale by a factor $(1+z)$.  This leads to a
modification of Eq.\ (2) such that

\begin{equation}
\tau \rightarrow \frac{\tau}{1+z}
\end{equation}

\noindent
The two panels of Fig.~\ref{fig7} each show three Fourier power spectra
constructed from the same blue and red passband light curves for
Fig.~\ref{fig6}, but with the modifications to Eq.\ (2) given in
Eq.\ (4).  The expected effect of extending the power spectrum to
higher temporal frequencies is clearly visible, and although an
approximate power law shape is retained the shape at low frequencies
now shows a noticeable departure from a power law (linear)
relationship.  In order to measure timescales, the same procedure was
carried out as for the data in Fig.~\ref{fig6}, involving fitting the
function in Eq.\ (3).  The best fit values for the blue and red
passbands were $9.7 \pm 0.1$ and $11.4 \pm 0.6$ years respectively.
These values are in good agreement, and well below the timespan of the
data giving confidence that a real timescale is being measured.

\begin{figure*}[t]
\setlength{\unitlength}{1mm}
\begin{picture} (200,45) (0,45)
\includegraphics[width=0.5\textwidth]{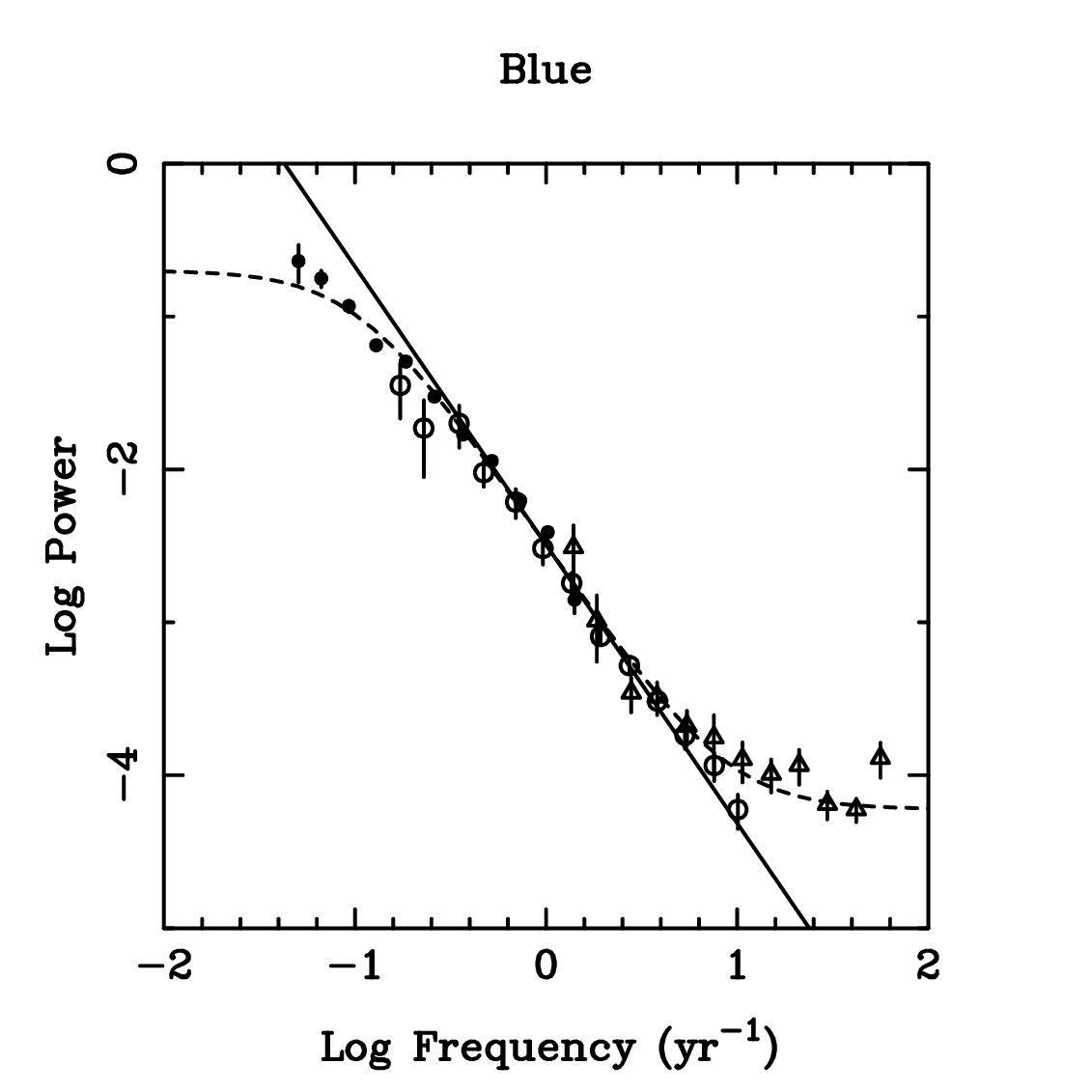}
\end{picture}
\begin{picture} (200,45) (-90,0)
\includegraphics[width=0.5\textwidth]{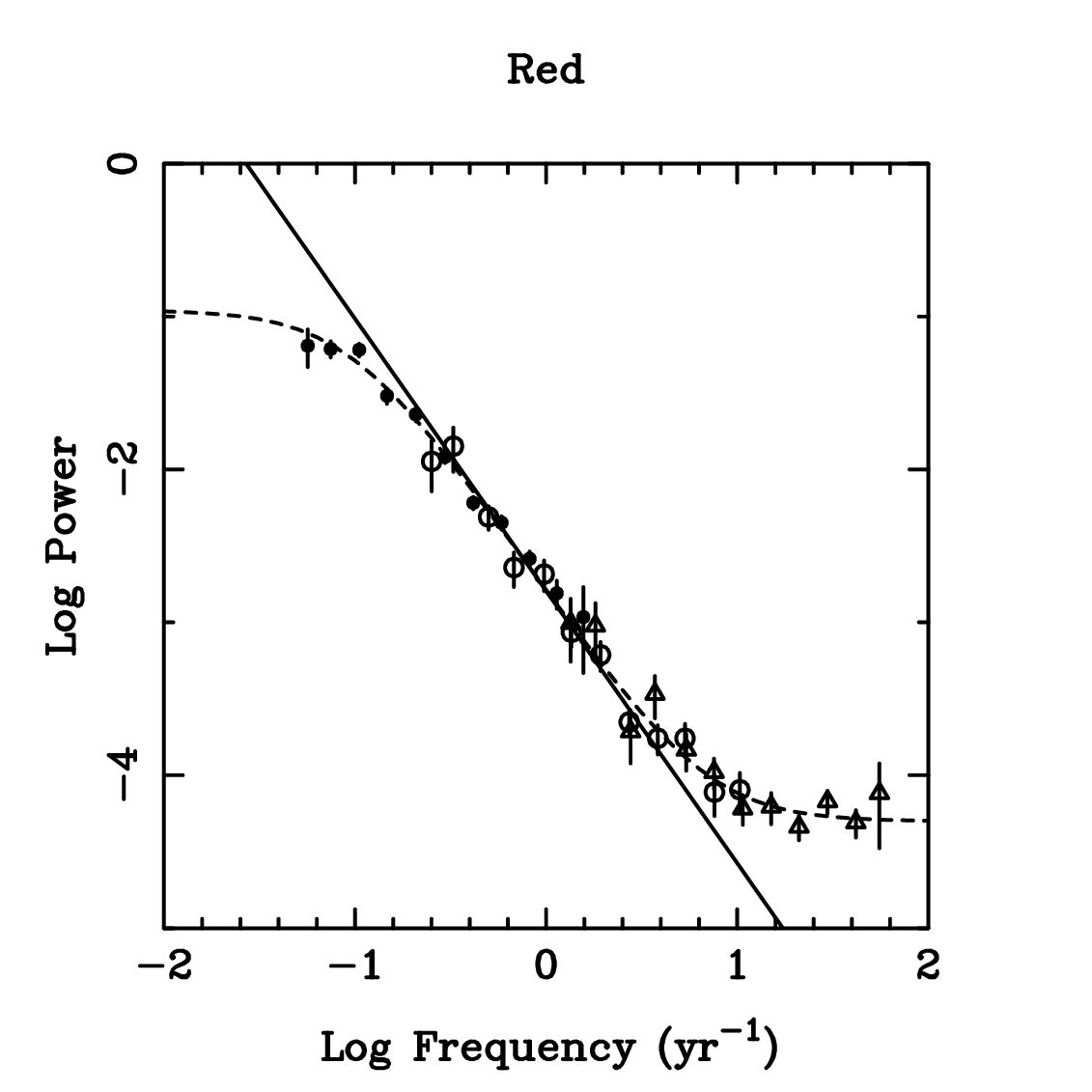}
\end{picture}
\caption{Fourier power spectra in the quasar rest frame for the data in
 Fig.~\ref{fig6}.  The best fit curves to the data are also shown as
 for Fig.~\ref{fig6}.
 \label{fig7}}
\end{figure*}

\section{Accretion disc structure}

Until recently the size of AGN accretion discs was not known with any
degree of certainty.  We first note that the accretion radius $r_{acc}$
given by

\begin{equation}
r_{acc} = \frac{2GM}{c_{\infty}^2}
\end{equation}

\noindent
where $c_{\infty}$ is the sound speed at infinity is an absolute upper
limit for the size of an accretion disc around a black hole of mass
$M$.  For a quasar this is a huge distance, around 30 parsecs, and not
restrictive as an upper limit.  A much more stringent limit is
derived by Goodman (2003), who uses the Toomre instability parameter
for Keplerian motion

\begin{equation}
Q \approx \frac{\Omega^{2}}{2 \pi G \rho}
\end{equation}

\noindent
to derive a radius $r_{Q=1}$ beyond which $Q < 1$ and local
gravitational instability occurs, and hence self-gravity terminates
the disc structure.  For typical parameter values this occurs at 
$r_{Q=1} \sim 10^{-2}$ pc.  Goodman then goes on to examine various
ways of creating larger quasar discs, but concludes that they probably
do not extend beyond $10^{-2}$ pc or 1000 Schwarzschild radii.

An alternative approach to determining the size of accretion discs is
discussed by Lyubarskii (1997).  In this paper he makes the point that
fluctuations in the accretion rate lead to flicker noise, and a power
law Fourier power spectrum, with a cut-off related to the radius of the
outer edge of the accretion disc.  In his model, long timescale
fluctuations are generated in the outer part of the accretion disc
and propagate inwards where they cause changes in brightness in the
optically emitting region near the disc centre.  The outer radius
$r_{out}$ of the accretion disc is given by Lyubarskii as

\begin{equation}
r_{out} = (\alpha \sqrt{G M} f_{min}^{-1})^{2/3}
\end{equation}

\noindent
where $\alpha$ is the viscosity parameter and $f_{min}$ is the
frequency of the break in the power law spectrum of variations.
King et al.\ (2004) develop this scheme further with an explicit
physical model for disc variability, invoking local dynamo processes
driving angular momentum loss through an outflow wind.

The frequency $f_{min}$ in Eq.\ (7), as defined by Lyubarskii (1997),
is basically the same frequency $f_{c}$ from Eq.\ (3) measured as the
break in the power law in Fig.~\ref{fig7}.  If we substitute the rest
frame value from above of about 11 years for $f_{min}^{-1}$ in
Eq.\ (7), and adopt $\alpha = 0.1$ (\cite{b98,g03a}) we obtain
$r_{out} = 10^{-2}$ pc or $3 \times 10^{16}$ cm.

\section{Microlens masses}

In the event that the quasar variations are dominated by the effects of
microlensing, it is appropriate to use the light curves measured in the
observer's frame, and the cut-off frequency $f_{c}$ from Eq.\ (3) is a
measure of the characteristic mass of the microlenses.  To see this we
note that the distance over which the gravitational field of a lensing
object has a significant effect on the image, known as the Einstein
radius $\theta_{e}$, and defined by

\begin{equation}
\theta_{e} = \left\{\frac{4Gm}{c^{2}R}\right\}^{\frac{1}{2}}
\end{equation}

\noindent
varies as the square root of the lens mass $m$.  In this equation $R$
is the distance to the lens and $G$ and $c$ are the gravitational
constant and speed of light respectively. If the lensing object is
moving across the line of sight with tangential velocity $v_{t}$,
the image will brighten and fade on the timescale that the source lies
within the Einstein radius.  Hence the characteristic timescale of
variation $t_{c}$ will be given by

\begin{equation}
t_{c} = \frac{R\theta_{e}}{v_{t}}
\end{equation}

\noindent
and combining Eq.\ (8) and Eq.\ (9)

\begin{equation}
m = \frac{c^{2}}{2G} \frac{v_{t}^{2}t_{c}^{2}}{R}
\end{equation}

\noindent
Thus $m$ varies as $t_{c}^{2}$.  If we adopt a characteristic value of
600 km s$^{-1}$ for $v_{t}$ and assume that most lenses lie at a
redshift $z \sim 0.5$ (\cite{t84}) then by using the lower limit to
the timescale of 24 years measured from the data in Fig.~\ref{fig6}, we
can obtain a value for the minimum microlens mass $m \sim 0.4 M_\odot$.

\section{Discussion}

After the first primitive estimates of timescale discussed in the
Introduction, and when the Fourier power spectrum was plotted in the
form of a log/log plot, it was clear that there was no obvious break in
a linear relation.  This situation persisted as more data was added,
and it was in this context that the short term MACHO data was added as
the only realistic way of extending the baseline.  It is quite
remarkable that for the case of the observer's reference frame the
straight line relationship was maintained, covering three orders of
magnitude in time and reinforcing the result of Hawkins (2001).

Until recently, little attention had been paid in this project to an
analysis of the light curves in the quasar rest frame.  This was
largely due to the author's interest in the microlensing interpretation
of quasar variability where the observer's frame is more appropriate.
It came therefore as something of a surprise to find a clearly
measurable break in the power spectrum for data analysed in the quasar
frame.  This meant that a definite measure of timescale could be made
rather than a lower limit, and has interesting implications for quasar
accretion discs.

The figure for the outer radius of the accretion disc of 
$3 \times 10^{16} cm$ refers to the entire disc out to the point where
graviational instability sets in.  It is interesting to compare this
with measurements of disc size from microlensing in multiply lensed
quasar systems.  In this situation the radius which is measured is that
of the luminous part of the disc.  For the system Q0957+561, Pelt et
al. (1998) find for the source radius $r_{s} \sim 3 \times 10^{15} cm$,
and for Q2237+0305 Kochanek (2004) also finds
$r_{s} \sim 3 \times 10^{15} cm$.  The optically luminous part of the
disc would thus appear to be about a tenth of the diameter of the whole
disc.

If we assume that the observed variations are intrinsic to the quasar
then the coincidence between $r_{Q=1}$ and $r_{out}$ is remarkable,
given the completely different bases on which they were calculated.
It is interesting to note that $10^{-2}$ pc or 10 light days is
also the canonical value for the size of the broad line region in AGN
(\cite{p99}).  It is tempting to equate the break up of the outer edge
of the accretion disc due to gravitational instability with the cloud
structure and turbulence of the broad line region.  Should this
measurement of the size of the accretion disc prove robust, it will
provide a fundamental constraint on the structure of the unified model,
and facilitate the determination of other model parameters.  However,
before this measurement can be accepted with confidence, there are
still some issues which need to be addressed.

The small size of the accretion disc presents some difficulties,
which have been discussed by Goodman (2003) in relation to his own
work.  The main problem is to find a way of replenishing the disc with
gas of small angular momentum at a high enough rate to build a
super-massive black hole. His suggestion that it might come from a
low angular momentum tail of gas from the bulge seems quite plausible.

To make full use of the results reported here it is most important
that Fourier power spectra of synthetic light curves for various
accretion disc models are made available in the literature.  Typically,
power spectra take the form of a power law rise to low frequencies,
terminated by a break corresponding to some characteristic timescale.
Free parameters include the power law index which reflects the nature
of the accretion process, the break frequency which is a measure of the
size of the accretion disc, and the amplitude which depends on the
mass diffusion rate.  A good start towards producing testable models
has already been made (\cite{k98,k03}), and hopefully there is more to
come.

It is interesting to note that the power law index measured for optical
variations in this paper ($\alpha = -1.77$) is similar to that found for
X-ray variations.  For example, Edelson \& Nandra (1999) find
$\alpha = -1.74$ for NGC 3516, and Uttley et al. (2002) obtain similar
results for a small sample of AGN.  However, the timescales vary by two
orders of magnitude between optical and X-ray, and it is clear that
different things are being measured in the two passbands.
Understanding the relationship between the two provides an interesting
challenge.

In the event that the observed variation is dominated by the effects of
microlensing, the lower limit of $m \sim 0.4 M_\odot$ for the
microlensing bodies is considerably larger than earlier estimates using
this approach (\cite{h96}).  Such objects would lie in the same mass
range as those detected by the MACHO project (\cite{a97}), although
they would have to be more abundant to be responsible for the extent of
variability observed.  Estimates of lens mass from the study of
microlensing events in multiply lensed quasar systems tend to give
upper limits somewhat similar to the lower limit quoted above for the
present work.  For example, Kochanek (2004) finds an upper limit to the
mean of the mass function of microlensing bodies in the system
Q2237+0305 to be around $m \leq 0.2 M_{\odot}$, and Refsdal et al.
(2000) find an upper limit to the mass of microlensing bodies in the
quasar system Q0957+561 of $m \leq 0.5 M_{\odot}$.

\section{Conclusions}

In this paper we have combined datasets from two AGN monitoring
programmes to construct Fourier power spectra in blue and red passbands
on timescales from 10 days to 25 years.  We have investigated different
methods of measuring timescales of variation in AGN light curves, and
concluded that Fourier power spectrum analysis provides the best
approach.  We have used the combined datasets to construct Fourier
power spectra covering nearly three orders of magnitude in time in both
the observer's frame and the quasar rest frame.  Timescales were then
measured by fitting a function designed to detect a turn over in the
power spectrum.

If the variations are interpreted as resulting from gravitational
microlensing then it is appropriate to use the observer's frame to
measure timescales.  In the observer's frame the power spectrum was
found to have an approximately power law shape with no significant
break, and a lower limit of 24 years was obtained for the timescale of
variation.  This corresponds to a minimum mass for microlensing bodies
of $m \sim 0.4 M_\odot$.  This mass is larger than that found in
earlier studies of quasar variation, and typical of the mass of bodies
found in the Galactic halo by the MACHO project.

If the variations are seen as intrinsic to the AGN then the effect of
time dilation must be corrected for.  This results in a modified
Fourier power spectrum which retains its power law shape on short
timescales, but with a break towards higher frequencies.  This break
was shown to correspond to 11 years in the quasar rest frame.  We
interpret this as a characteristic timescale for variations intrinsic
to the AGN.  This timescale is used to estimate the size of an AGN
accretion disc which we find to be $10^{-2}$ pc or about 10 light days.
This is consistent with the radius at which the onset of self-gravity
may terminate the disc, and roughly coincident with the position of the
broad line region.  This suggests the possibility that the broad line
region is associated with the break up of the outer part of the
accretion disc.

\end{document}